\documentclass[lotsofwhite,charterfonts]{patmorin}
\usepackage{graphicx,graphics,epsfig,amsfonts,amsthm,amsmath}
\usepackage{amsthm}

\newcommand{\comment}[1]{}

\newcommand{\seclabel}[1]{\label{sec:#1}}
\newcommand{\Secref}[1]{Section~\ref{sec:#1}}
\newcommand{\secref}[1]{\mbox{Section~\ref{sec:#1}}}

\newcommand{\figlabel}[1]{\label{fig:#1}}
\newcommand{\Figref}[1]{Figure~\ref{fig:#1}}
\newcommand{\figref}[1]{\mbox{Figure~\ref{fig:#1}}}

\newcommand{\eqlabel}[1]{\label{eq:#1}}

\newtheorem{thm}{Theorem}{\bfseries}{\itshape}
\newcommand{\thmlabel}[1]{\label{thm:#1}}
\newcommand{\thmref}[1]{Theorem~\ref{thm:#1}}

\newtheorem{lem}{Lemma}{\bfseries}{\itshape}
\newcommand{\lemlabel}[1]{\label{lem:#1}}
\newcommand{\lemref}[1]{Lemma~\ref{lem:#1}}

{\bfseries}{\itshape}

{\bfseries}{\itshape}

{\bfseries}{\itshape}

\newtheorem{assumption}{Assumption}{\bfseries}{\rm}

\newcommand{\R}{\mathbb{R}}

\usepackage{marvosym}

\newcommand{\RR}{\mathbb{R}}
\newcommand{\D}{\mathbb{D}}
\newcommand{\PROB}{\Pr}
\newcommand{\EXP}{\mathrm{E}}

\title{\MakeUppercase{On the Expected Maximum Degree 
	of Gabriel and Yao Graphs}}
\author{Luc Devroye
	\and Joachim Gudmundsson
	\and Pat Morin}

\begin{document}
\maketitle
\begin{abstract}
Motivated by applications of Gabriel graphs and Yao graphs in wireless
ad-hoc networks, we show that the maximal degree of a random Gabriel
graph or Yao graph defined on $n$ points drawn uniformly at random from
a unit square  grows as $\Theta ( \log n / \log \log n)$ in probability.
\end{abstract}

\section{Introduction}

Wireless ad-hoc networks consist of computers (or sensors) capable of
communicating wirelessly with each other without any centralized
information, infrastructure, or organization.  A common mathematical model
of such networks is the \emph{unit disk graph} in which the nodes consist
of $n$ points in $\R^2$ and an edge exists between two nodes if and only if
the distance between them is at most $r$.  Depending on the value of $r$,
which represents the transmission range of the wireless transmitters, the
network can be anything ranging from a set of isolated vertices to the
complete graph.

The lack of centralized management and organization that occurs in ad-hoc
networks means that individual nodes in the network typically only have
local information about the nodes that they can communicate directly with.
This makes even basic tasks, such as routing, highly non-trivial because
the combination of complete lack of organization and the unit disk graph
topology is too unwieldy.

One approach to taming ad-hoc networks has been to compute the
intersection of the unit disk graph with some ``nice'' proximity graph.
If the right proximity graph is chosen, the resulting graph will
remain connected (if the original unit disk graph is connected) and will
inherit some of the nice properties of the proximity graph.   Ideally, the
intersection can be computed locally, so that individual nodes can locally
determine which of their incident edges belong to the intersection.

One such approach computes the intersection of the unit disk graph with
the Gabriel graph \cite{gs69}.  The Gabriel graph contain an edge between
two points $u$ and $v$ if and only if the disk whose diameter is $uv$
contains no points other than $u$ and $v$ (see \figref{gabrieldef}).  The
Gabriel graph is planar and therefore has only a linear number of edges.
Algorithms for routing on planar graphs can be applied to the resulting
graph or, more commonly, these algorithms can be used for recovery when
routing heuristics fail.  A number of routing algorithms and protocols
have been proposed based on this strategy \cite{bfno03,bmsu01,kk00}.

\begin{figure}
\begin{center}{\includegraphics[scale=0.7]{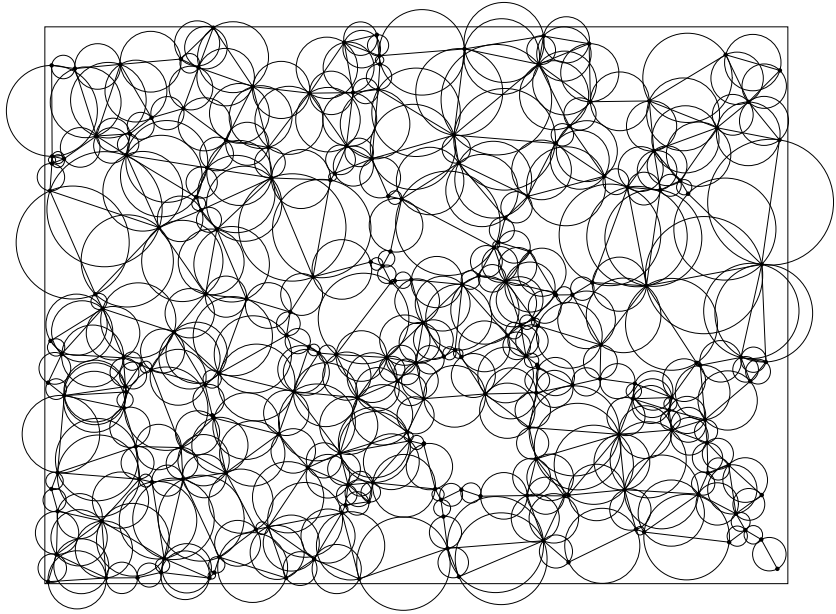}}\end{center}
\caption{A point set with its Gabriel graph. No circle has
any data point in its interior and every circle has an edge
as its diameter.}
\figlabel{gabrieldef}
\end{figure}

Another suggested approach uses the Yao graph \cite{y82}. Refer to
\figref{yaodef}. Let $p$ be a positive integer, let $\theta=2\pi/p$,
and let $u$ be a point in $\R^2$.  The $i$-cone of $u$ is the set of all
points $w\in\R^2$ such that $\angle quw \in [(i-1)\theta,i\theta)$, where
$q = u+(1,0)$.  The \emph{$\theta$-Yao graph} contains an edge from $u$
to the nearest point in each of $u$'s $i$-cones, for $i=1,\ldots,p$.
For any constant $p\ge 6$, the $\theta$-Yao graph has at most $pn$
edges and is a \emph{spanner}; for any two vertices $u$ and $v$, the
$\theta$-Yao graph contains a path whose Euclidean length is at most
$t\cdot\|uv\|$, where $\|uv\|$ denotes the Euclidean distance between $u$
and $v$ and $t=1/(1-2\sin(\theta/2))$ is called the \emph{stretch factor}.
When applied in the context of unit disk graphs, if there is a path of
Euclidean length $\|uv\|_U$ in the original unit disk graph, then there
is a path of length at most $t\cdot\|uv\|_U$ in the intersection of
the unit disk graph and the $\theta$-Yao graph.  Routing strategies
based on the Yao graph attempt to find power-efficient routing paths
\cite{glsv02,lwy01,wlbw01,svz07}.

\begin{figure}
\begin{center}{\includegraphics{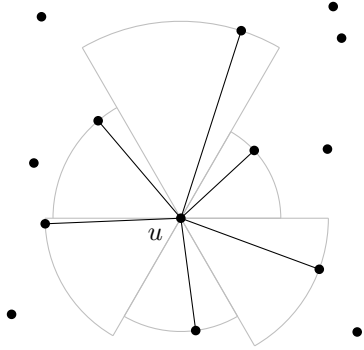}}\end{center}
\caption{The edges defined by a node $u$ in a $(\pi/3)$-Yao graph.}
\figlabel{yaodef}
\end{figure}

\subsection{New Results}

Motivated by the above applications in wireless networks, the current paper
studies the Gabriel graph and Yao graph of $n$ points uniformly and
independently distributed in a unit square.  This distribution assumption
can be used to approximately model the unorganized nature of ad-hoc
networks and is commonly used in simulations of such networks \cite{tma09}.
Additionally, some types of sensor networks, especially with military
applications, are specifically designed to be deployed by randomly placing
(scattering) them in the deployment area. This distribution assumption
models these applications very well.

We show that the maximum degree of any node in a Gabriel
graph or a Yao graph is concentrated at $\Theta(\log n/(\log\log
n))$.\footnote{Throughout this paper $\log x$ denote the natural logarithm
of $x$.} More specifically, if $\Delta$ is the maximum degree of either
graph, then we show that there exists constants $a$ and $b$, such that
\[
    \lim_{n\rightarrow\infty}
       \PROB\left\{\Delta\in \left[\frac{a\log n}{\log\log n},
                         \frac{b\log n}{\log\log n}\right]\right\} = 1
       \enspace .
\]
For Gabriel graphs, we show this for $(a,b)=(1/12,1)$ and for Yao graphs
we show it for $(a,b)=(1/8,4)$.  The maximum degree is particularly
important in wireless networks, since the degree of a node directly impacts
the amount of bookkeeping the node must do.  With wireless nodes typically
being battery operated and often memory- and computation-constrained, the
degree of a node should hopefully be as small as possible in order to
minimize this bookkeeping.

\subsection{Related Work}

A \emph{random Gabriel graph} in this paper is a Gabriel graph for $n$
points drawn uniformly and at random from $[0,1]^d$.  Its key properties
were studied in great depth by Matula and Sokal \cite{ms80}.  For example,
the expected number of edges grows as $2^{d-1} n$ \cite{d88,ms80}.  The
length of an edge taken at random from all edges has expected value and
standard deviation $\Theta(n^{-1/d})$ \cite{d88}.  These properties hold
also for many non-uniform distributions \cite{d88}.  

For a uniform Poisson process, introduced to avoid edge effects,
Bern, Eppstein and Yao \cite{bey91} showed that the expected value of
the maximal degree of a Delaunay triangulation grows as $\Theta(\log
n/\log\log n)$.  For that model, their proof also works for Gabriel
graphs.  It is known that the Gabriel graph is a subgraph of the Delaunay
triangulation (see Toussaint \cite{t80b}), so that our upper bound would
in fact follow without too much work from the cited paper.  Our work on
Gabriel graphs differs in three aspects:

\begin{enumerate}
\item We show convergence in probability: the fact that the expected
maximal degree grows as $\Theta(\log n /\log \log n)$ does not imply that the
probability of obtaining such large maximal degrees tends to one. We
show it does.

\item We deal with a fixed sample size model on a compact set, not the
Poisson model on the entire plane.

\item Our proofs are different.
\end{enumerate}

The relative neighborhood graph is obtained by joining all pairs whose
loon is empty, where the loon defined by a pair is the intersection of two
spheres of equal radius, each having one point as center and the other
point on its surface (see Toussaint \cite{t80a}).  As it is a subgraph of
the Gabriel graph, our results imply that its maximal degree is $O(\log
n /\log \log n)$ in probability.  For a general discussion of proximity
graphs and their applications, we refer to the survey papers by Toussaint
\cite{t80b,t82}.  For an application of the relative neighbourhood graph to
wireless networks, see Karp and Kung \cite{kk00}.

To the best of our knowledge, random Yao graphs have not been studied
previously. Although researchers have been interested in spanners having
small maximum degree (see, the textbook by Narasimhan and Smid \cite{ns07} for
a survey), most research in this area has been on constructing
spanners that have low degree in the worst-case.  Some of these
constructions have been adapted for use in the unit disk graph model of
wireless networks \cite{wl06}, but the computation of these spanners is not
quite as straightforward and local as that of Yao graphs.

The remainder of this paper is organized as follows.  \Secref{gabriel}
presents our results on Gabriel graphs.  \Secref{yao} presents our
results on Yao graphs.  Each of these sections concludes with a summary
and discussion of possible generalizations and limitations.

\section{Gabriel Graphs}
\seclabel{gabriel}

In this section, we prove bounds on the maximum degree of vertices in a
Gabriel graph.  Before we begin, we discuss an equation that is central to
all our upper and lower bound, as well as many other bounds of this type.

Let $c>0$ be a constant, and let $k=c\log n/\log\log n$.  In all our
bounds, the value $k^k$ appears at some point in the computations.  Note that
\begin{equation}
    k^k = n^{c\left(1+\frac{\log c - \log\log\log n}{\log\log n}\right)} = n^{c-o(1)}
   \enspace . \eqlabel{central}
\end{equation}
In particular, $k=O(n^c)$ and, for any $\epsilon >0$,
$k=\Omega(n^{c-\epsilon})$.

\subsection{A lower bound}
\seclabel{gabriel-lower-bound}

In this section, we prove the following result.

\begin{thm}\thmlabel{gabriel-lower-bound}
For a random Gabriel graph defined on $n$ points drawn
independently from the uniform distribution on
$[0,1]^2$, 
\[
\lim_{n\to\infty} \PROB \left\{ \hbox{\rm maximal degree}~< {c \log n \over
\log \log n} \right\} = 0
\]
for all $c < 1/12$.
\end{thm}

\begin{proof}
We start with a technical construction of a region
and then a point configuration.
Given an integer $k$ and positive number $r$.
Define the angle $\xi = 2\pi / (3k)$,
and partition the plane into $3k$ sectors
of angle $\xi$ each, with center at the origin.
We refer to \figref{pearldef} for further explanations.

\begin{figure}[htbp]
\begin{center}\includegraphics{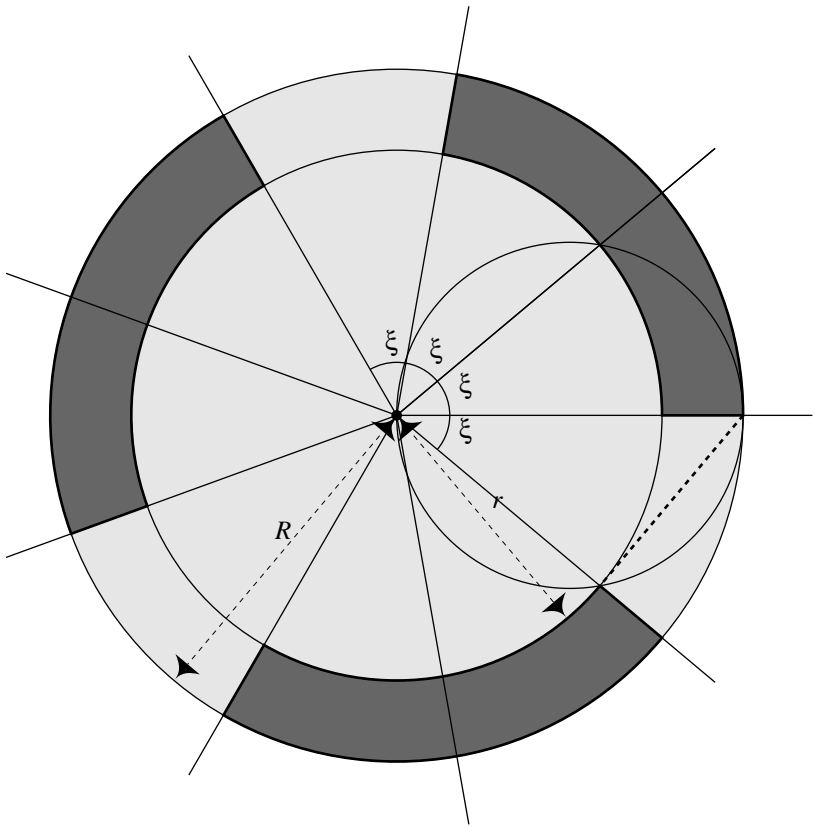}\end{center}
\caption{The definition of a pearl}
\figlabel{pearldef}
\end{figure}

Draw two concentric circles of radii $r$ and $R$
with $R > r$, so that $r = R \cos \xi$.
If the sectors are numbered $C_1, C_2, \ldots, C_{3k}$
(clockwise)
and the circles are $S_r$ and $S_R$, then we
mark $k$ regions (shown in darker color in \figref{pearldef}). These regions are of the form
$(S_R - S_r) \cap (C_{3i+1} \cup C_{3i+2})$ for
$0 \le i \le k-1$. 
Call these regions pearl regions
and denote them by $P_1 , \ldots, P_k$.
Any circle with
its diameter being the segment linking the origin with
any point in a pearl region totally avoids any other
pearl region. To see this, refer to \figref{pearldef}
and recall that $r = R \cos \xi$.

Assume we are given $m$ points in the plane,
$x_1,\ldots,x_m$ and a center $x$.
If $x+A$ denotes the translate of a set $A$ by $x$,
then we call $x$ a {\it tiara} for $x_1 , \ldots , x_m$
if exactly $k$ of the points $x_i$ fall in $x+S_R$,
and if each set $x+P_j$ covers exactly one of
these $x_i$'s.
If we construct the Gabriel graph for $x,x_1,\ldots,x_m$,
then the degree of the vertex at $x$ is at least $k$
if $x$ is a $(k,r)$ tiara for $x_1,\ldots,x_m$.

The above construction and definitions are
for any point sets.
Assume that a random sample of size $n$ is drawn
from the uniform distribution on $[0,1]^2$,
and denote it by $X_1 , \ldots, X_n$.
Define $k = \max( 3, \lfloor c \log n / \log \log n \rfloor)$
and $r = 1/\sqrt{n}$.
We say that $X_i$ is a {\it jewel}
if $X_i$ is a $(k,r)$ tiara for $\{ X_j : j \not= i \}$
and if $X_i$ is at distance at least $2r$ from the
perimeter of $[0,1]^2$. Note that $R \le 2r$, so that
$X_i + P_j \subseteq [0,1]^2$ for all $j$.

We compute the probability that $X_1$ is a jewel
given $X_1 = x$, provided that $x$ is at distance at
least $2r$ from the perimeter of the unit square.
Note that this probability may be written as
a multinomial probability. If $p$ is the area
of $x+P_j$, we have in particular,
\[
\begin{aligned}
\PROB \{ X_1 ~\hbox{\rm is a jewel} | X_1 = x \}
&=  {(n-1)! \over (n-1-k)!} \prod_{j=1}^k p \times \left( 1 - \pi R^2 \right)^{n-1-k} \\
&=  {(n-1)! \over (n-1-k)!} \prod_{j=1}^k p \times \left( 1 - \pi/n\cos^2\xi \right)^{n-1-k} \\
&\ge  {(n-1)! \over (n-1-k)!} \prod_{j=1}^k p \times \left( 1 - 2\pi/n \right)^{n-1-k} & \mbox{[since $\xi \le 2\pi/9 < \pi/4$]} \\
 &\ge  (n-k)^k p^k (1-2\pi/n)^n \\
&\ge  (n-k)^k p^k (1/3)^{2\pi} \enspace . 
\end{aligned}
\]
As $k \ge 3$, we have $\xi \le 2\pi/9 < 1$, so that
$\tan \xi = \sin \xi / \cos \xi \ge \xi$.
Therefore, 
\[
p = (R^2 - r^2) \xi = R^2 \xi \sin^2 \xi
  = r^2 \xi \tan^2 \xi \ge \xi^3/n \enspace .
\]
Resubstitution yields
\[
\begin{aligned}
\PROB \{ X_1 ~\hbox{\rm is a jewel} | X_1 = x \}
& \ge (n-k)^k {\xi^{3k} \over n^k} \left( {1 \over 3} \right)^{2\pi} \\
& =   \left( 1 - {k \over n} \right)^k  
    \left( {1 \over 3} \right)^{2\pi}
    \xi^{3k} \\
& = \left( 1 - {k \over n} \right)^k  
    \left( {1 \over 3} \right)^{2\pi}
    \left({2\pi\over 3}\right)^{3k} k^{-3k}\\
& \ge k^{-3k}
\end{aligned}
\]
when $n$ is large enough, uniformly over all $x$
at distance at least $2r$ from the perimeter of the unit square.
We may now uncondition. If $N$ is the 
number of jewels among the data points, we have
\[
\EXP[N]
 = n \PROB \{ X_1 ~\hbox{\rm is a jewel} \}
 \ge n (1-4r)^2 k^{-3k}
 \sim n k^{-3k}~.
\]
If $k$ is as we picked it, and $c < 1/3$,
then $\EXP[N] \to \infty$.
This is not quite enough to show that 
$\PROB \{ N > 0 \} \to 1$.
There are several routes one can follow at this
point: one could Poissonize the sample
size; one might redefine jewels so that
at most one jewel occurs in any region of
a regular grid. Both tricks create
enough independence to get by.
Instead, we opt to use the second moment
method (for references, see Palmer (1985) or
Alon, Spencer and Erd\"os (1992)). When
applied to a counting random variable $N = \sum_{i=1}^n Y_i$,
where the $Y_i$'s are $\{ 0,1 \}$-valued with
a permutation-invariant joint distribution,
the second moment method implies that
$N/\EXP[N] \to 1$ in probability whenever
$\EXP[N]\to \infty$ and
\[
\limsup_{n\to\infty} \frac{\EXP[Y_1 Y_2]}{\EXP[Y_1]\EXP[Y_2]} \le 1 \enspace .
\]
In our case, we only need to verify the latter
condition when $Y_i$ is the indicator that
$X_i$ is a jewel, so that $N$ is the number of jewels.
Let $A$ be the event that $X_1$ or $X_2$ is within
$2r$ of the perimeter of the unit square, or that
$\| X_1 - X_2 \| \le 4r$.
On $A^c$, the complement of $A$, we have, by
the multinomial argument given above, but now
applied to two tiaras,
\[
\EXP[Y_1 Y_2 | A^c]
= { (n-2)! \over (n-2-2k)! } p^{2k} (1-2\pi R^2)^{n-2-2k}~,
\]
where $p$ is the area of a pearl region $P_j$.
We recall that
\[
\EXP[Y_1]\ge
{(n-1)! \over (n-1-k)!} p^k \left( 1 - \pi R^2 \right)^{n-1-k}
\ge k^{-3k}
\]
for $n$ large enough.
Thus, for such large $n$,
\[
\begin{aligned}
{\EXP[Y_1 Y_2] \over \EXP[Y_1]\EXP[Y_2]} 
&=
{\PROB\{A\}\EXP[Y_1 Y_2 |A ] \over \EXP[Y_1]\EXP[Y_2]}
+ {\PROB\{A^c\}\EXP [ Y_1 Y_2 | {A^c} ] \over \EXP[Y_1]\EXP[Y_2]} \cr
&\le
{\PROB \{ A \} \over \EXP[Y_1]\EXP[Y_2]}
+ {\EXP [ Y_1 Y_2 | A^c ] \over \EXP[Y_1]\EXP[Y_2]} \cr
&\le
{8r + 16 \pi r^2 \over \EXP[Y_1]\EXP[Y_2]}
+ { (n-2)! \over (n-2-2k)! } p^{2k} (1-2\pi R^2)^{n-2-2k}\times
{(n-1-k)!^2 \over (n-1)!^2} p^{-2k} \left( 1 - \pi R^2 \right)^{2k+2-2n} \cr
&\le
{67 \over n^{1/2} k^{-6k}} + 1 + O(k^2/n) ~. \cr
\end{aligned}
\]
We are done if $n^{1/2} k^{-6k} \to \infty$. For this,
in the definition of $k$, we need only pick
$6c < 1/2$, or $c < 1/12$.
We have thus shown that $N/\EXP[N]\to 1$ in probability
when $c < 1/12$ in the definition of $k$.
We conclude that
$\PROB \{ N = 0 \}  \to 0$ for such choices of $c$.
Therefore,
\[
\lim_{n\to\infty}
\PROB \left\{ \hbox{\rm Maximal degree in Gabriel graph}
	   < {c \log n \over \log \log n } \right\}
	   = 0
\]
for all $c < 1/12$.
By Poissonization, or by
tighter bounding above, the constant $1/12$ can be improved.
\end{proof}

\begin{figure}
\begin{center}{\includegraphics{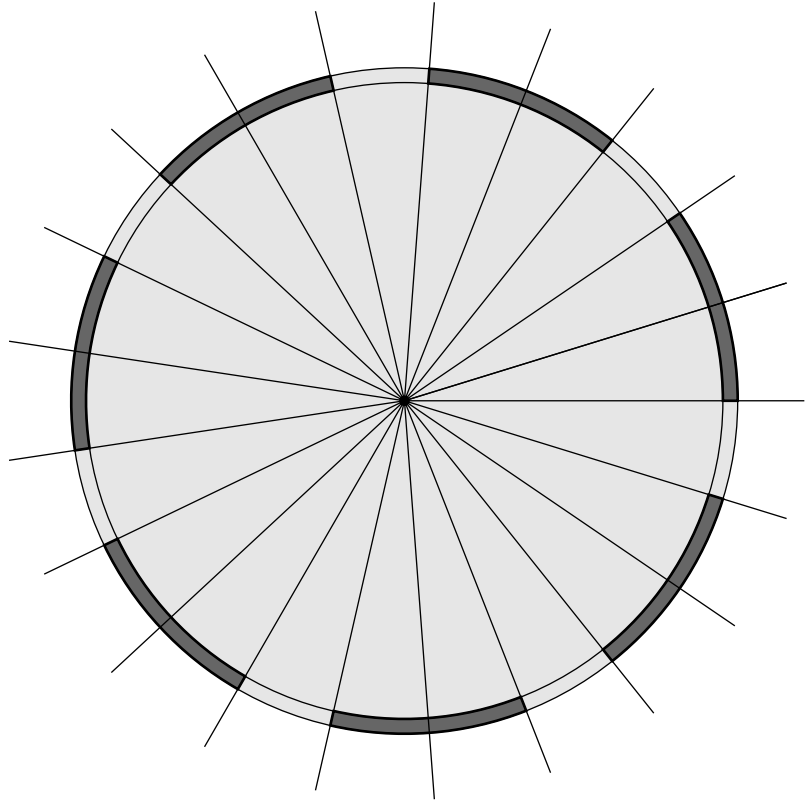}}\end{center}
\caption{The shaded regions define a $(7,r)$ tiara.}
\end{figure}

\subsection{An upper bound}
\seclabel{gabriel-upper-bound}

\begin{thm}\thmlabel{gabriel-upper-bound}
For a random Gabriel graph defined on $n$ points drawn
independently from the uniform distribution on
$[0,1]^2$, 
\[
\lim_{n\to\infty} \PROB \left\{ \hbox{\rm maximal degree}~> {c \log n \over
\log \log n} \right\} = 0
\]
for all $c > 1$.
\end{thm}

\begin{proof}
At a point $x$, partition the space into $k$ equal
sectors of angle $2\pi/k$ each, where $k = \lceil \sqrt{\log n} \rceil$.
Within each sector, we color the point nearest to $x$ 
red if its distance is less than $r = 3 \sqrt{\log n\, /n}$.
If a sector has a red point $y$, consider the perpendicular
line at $y$ to the segment $(y,x)$. Call this
line the separator. All points
in the same sector but at the same side as $x$ of the separator
are colored blue. In \figref{wedgedef}, these are precisely the points that
fall in the shaded wedge.
Finally, among all points, color those yellow that are
Gabriel neighbors and that are at least $r$ away from $x$.
We first claim that each Gabriel graph neighbor of $x$
is colored red, blue or yellow.
Indeed, any point $y$ excludes all points at the
other side of the separator---the side that does not contain $x$.
Thus, if there is a red point in the sector, only blue points
can possibly be Gabriel neighbors of $x$.

\begin{figure}
\begin{center}\includegraphics{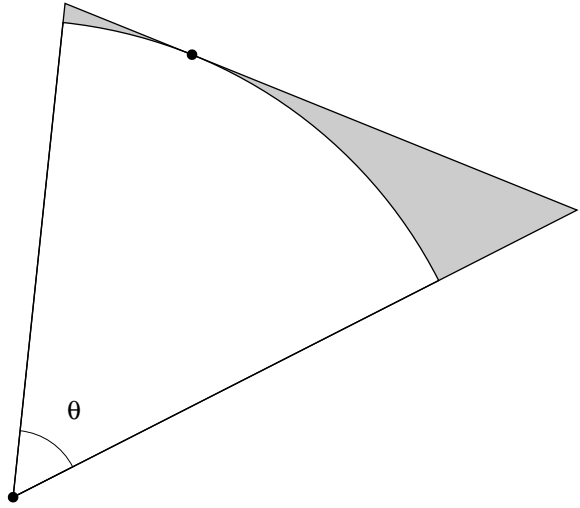}\end{center}
\caption{The definition of a wedge}
\figlabel{wedgedef}
\end{figure}

\Figref{gabriel-exclude} shows several points with their separators.
No point in the shaded area can be a Gabriel neighbor of the point at the
origin. Note that for every point in the shaded area, the Gabriel circle
through the origin contains another point.

\begin{figure}
\begin{center}\includegraphics[scale=0.5]{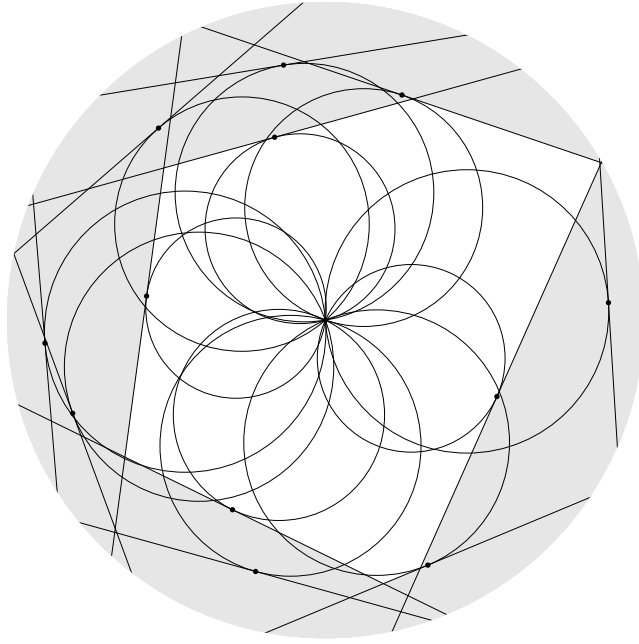}\end{center}
\caption{Several points and their separators}
\figlabel{gabriel-exclude}
\end{figure}

\Figref{wedge}
shows several sectors and red points, together
with the wedges in which blue points must
fall. The angle of each sector is $\theta = 2\pi / k$.

\begin{figure}
  \begin{center}
    \includegraphics{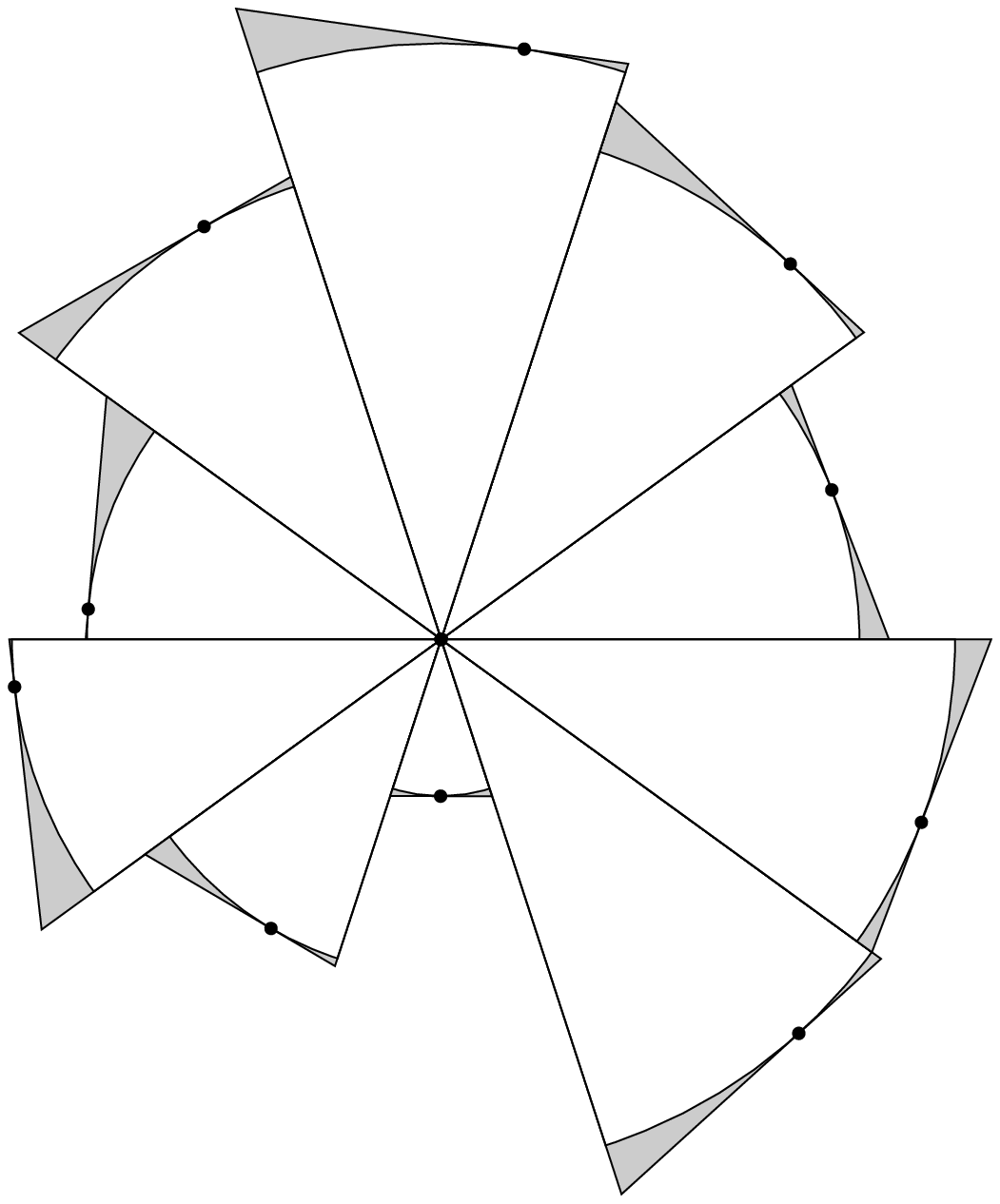}
  \end{center}
  \caption{Sectors, red points, and blue wedges.}
  \figlabel{wedge}
\end{figure}

Let $N_r , N_b, N_y$ be the total number of red, blue and yellow
points respectively.
Clearly, $N_r \le k = o(\log n /\log \log n)$.
Also, conditioning on $X_1 = x$,
\[
\begin{aligned}
\EXP [N_y | X_1 = x] 
&= n \PROB \{ X_2 ~\hbox{\rm is a Gabriel neighbor of}~ X_1 , \| X_2 - X_1 \| \ge r | X_1 = x \} \cr
&\le n (1-r^2/4)^{n-1} \cr
&\quad \hbox{\rm (because at least a $1/\pi$ fraction} \cr
&\quad \hbox{\rm of the Gabriel circle through $x$ and $X_2$ falls in the unit square)} \cr
&\le n e^{-9(n-1) \log n / (4 n)} \cr
&\sim n^{1-9/4} \cr
&\to 0~. \cr
\end{aligned}
\]
Thus, it suffices to study $N_b$.

As $\tan \theta \le \theta + \theta^3$ for $0 \le \theta \le 1$,
the area of each wedge is at most
\[
\begin{aligned}
{ r^2 \over 2} (\tan \theta - \theta)
&\le { 9 \log n \theta^3 \over 2n}  \cr
&\le { 72 \log n \pi^3 \over 2nk^3}  \cr
&\le { 1200 \log n \over n k^3} ~. \cr
\end{aligned}
\]
The total wedge area around $X_1 = x$ is thus not more
than
\[
{ 1200 \log n \over n k^2 } \le { 1200 \over n }~.
\]
Given $X_1 = x$ and the collection of red points,
the $n-1-N_r$ other points are uniformly distributed
on the unit square minus the $N_r$ circular
sectors just inside the wedges, and minus 
circular sectors of radius $r$ defined when no
red point is present in the sector. Call the density $f$
and its support set $S$. Clearly, 
$1 \le f 
\le 1/(1-\pi r^2)$.
Of the $n-1-N_r$ points, let $M$ denote the total
number of points falling in the wedges.
Clearly, $M$ is stochastically smaller
than a binomial random variable with parameters
$m = n-1-N_r$ and $p = 1200/n(1-\pi r^2)$.
In particular, using $l! \ge (l/e)^l$,
and letting $\PROB$ denote the conditional probability,
\[
\begin{aligned}
\PROB \{ M \ge l \}
&\le \sum_{j=l}^m {m \choose j} p^j (1-p)^{m-j} \cr
&\le \sum_{j=l}^\infty {(mp)^j \over j!}  \cr
&=   {(mp)^l \over l!} \sum_{j=0}^\infty {(mp)^j l! \over (l+j)!}  \cr
&\le {(mp)^l \over l!} \sum_{j=0}^\infty \left( {mp \over l} \right)^j  \cr
&=   {(mp)^l \over l! (1-mp/l)}  \cr
&\le   {(mpe/l)^l \over (1-mp/l)}  \cr
&\le   {(npe/l)^l \over (1-np/l)} ~. \cr
\end{aligned}
\]
We set $l = \lceil c \log n / \log \log n \rceil$ for a constant $c$
and note that $np/l = o(1)$.
By the union bound, 
the probability that for one of the $n$ data points,
the number of blue color points in the wedge collection
for that point is greater than or equal to $l$ does not
exceed 
\[
n \times {(npe/l)^l \over (1-np/l)} ~. 
\]
As $np = 1200+o(1)$, the above expression is for all $n$
large enough not more than
\[
2n (3600/l)^l~.
\]
This tends to zero when $c > 1$.

The probability that for one of the data points,
$N_y > 0$ is not more than
\[
n \times (1+o(1)) n^{1-9/4} \to 0~.
\]
Thus, we have shown that for $c > 1$, 
the probability that the maximal degree
exceeds $k+l$ tends to zero. As $k+l \sim l$, we are done.
\end{proof}

\subsection{Remarks}

\paragraph{Higher dimensions.}
Just as Bern, Eppstein and Yao (1991) showed for the
expected maximal degree in a Delaunay triangulation, 
the results for in probability convergence for Gabriel graphs
extend easily to $\RR^d$. In particular,
for any $d$, there exist constants $a>0$ and $b < \infty$ 
only depending upon $d$ such that
\[
\lim_{n\to\infty} \PROB \left\{ \hbox{\rm maximal degree}~\not\in 
  \left( {a \log n \over \log \log n} , { b \log n \over \log \log n} \right)
  \right\} = 0 \enspace .
\]

\paragraph{Edge lengths.}
The results on $N_y$ in the proof above
show that the expected number of Gabriel
edges of length at least $3 \sqrt{\log n \,/n}$
is $o(1)$. Hence, the probability that the maximal edge
length exceeds $3 \sqrt{\log n \,/n}$ tends to zero.
In $\RR^d$, the maximal edge length is
easily shown to be $O((\log n \,/n)^{1/d})$ in
probability.
In contrast, one can show that if $E_i$ is the
maximal edge length among the Gabriel edges incident to $X_i$,
then $(1/n) \sum_{i=1}^n \EXP[E_i]= O(n^{-1/d})$,
and that if 
$F_i$ is the
minimal edge length among the Gabriel edges incident to $X_i$,
then $(1/n) \sum_{i=1}^n \EXP[F_i]= \Omega (n^{-1/d})$. 

\section{Yao Graphs}
\seclabel{yao}

In this section we present our results on Yao graphs.  For simplicity
we consider $\theta$-Yao graphs with $\theta=\pi/2$.  The modifications
required for other (smaller) values of $\theta$ are discussed at the end
of this section.  The lower bound in \secref{yao-lower-bound} is obtained
using a construction and argument similar to the pearl used to prove
\thmref{gabriel-lower-bound}.  The upper bound in \secref{yao-upper-bound}
uses different arguments based on maxima.

For the upper bound, we change the distribution model slightly by rotating
it by $\pi/4$.  More precisely, let $\D^2$ denote the unit square rotated
by $\pi/4$.  The upper bound assumes that points are distributed uniformly
and independently in $\D^2$. At the end of this section, we discuss why
this slightly different assumption is necessary.

\subsection{A lower bound}
\seclabel{yao-lower-bound}

Our lower bound argument is similar to that used for Gabriel graphs, in
that we define a configuration of points whose existence implies a vertex
of degree $k$ and show that, with high probability, this configuration
exists in a random point set.  

\begin{thm}\thmlabel{yao-lower-bound}
For a random $\pi/2$-Yao graph defined by $n$ points drawn independently
from the uniform distribution on $[0,1]^2$,
\[
   \lim_{n\rightarrow\infty}
     \Pr\left\{\mbox{Maximal degree} < \frac{c\log n}{\log\log n} \right\} 
       = 0 \enspace ,
\]
for all $c < 1/8$.
\end{thm}

\begin{proof}
Refer to \figref{staircase}.a.  Let $r>0$ be a real number and let $k$ be
a positive integer.  Define $k$ square regions $P_1,\ldots,P_k$ where
$P_i = [(i-1)r/k,ir/k]\times [r-ir/k,r-(i-1)r/k]$.  These regions are
called \emph{steps}.

\begin{figure}
  \begin{center}
    \begin{tabular}{cc}
      \includegraphics{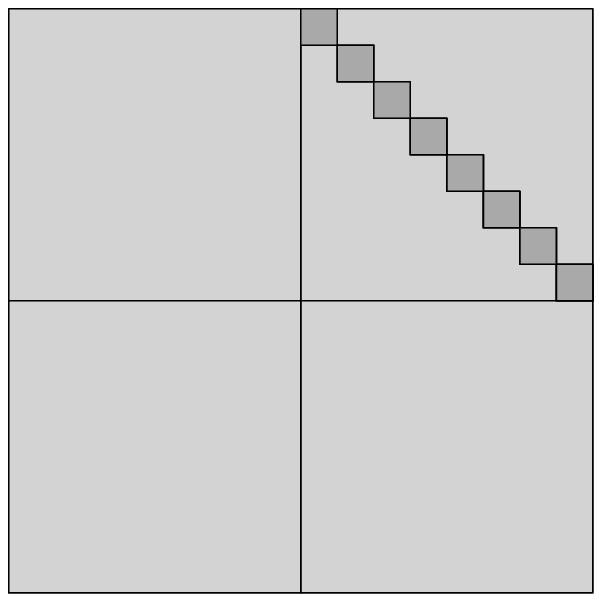} & \includegraphics{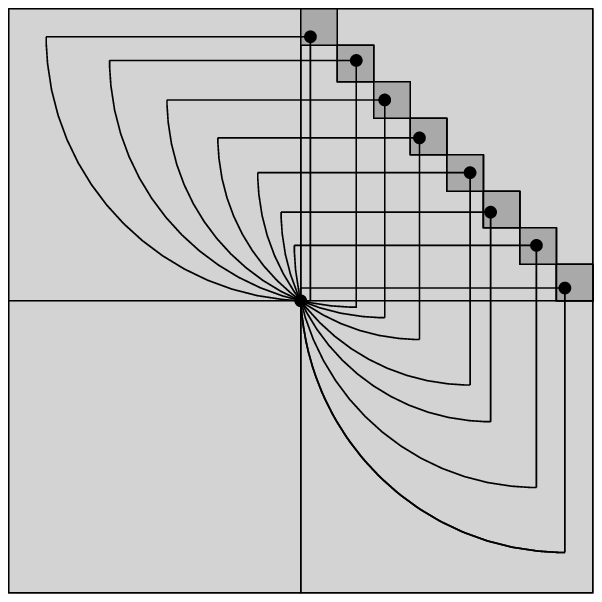} \\
      (a) & (b)
    \end{tabular}
  \end{center}
  \caption{An $(8,r)$ staircase.}
  \figlabel{staircase}
\end{figure}

Assume we are given $m$ points in the plane, $x_1,\ldots,x_m$ and a center
$x$.  Then we call $x$ a $(k,r)$-\emph{staircase} for $x_1 , \ldots , x_m$
if exactly $k$ of the points $x_i$ fall into the square $x+[-r,r]^2$ and if
each step $x+P_j$ covers exactly one of these $x_i$'s.
If we construct the $\pi/2$-Yao graph for $x,x_1,\ldots,x_m$ and $x$ is a
$(k,r)$-staircase for $x_1,\ldots,x_m$, then every point in each of the $k$
steps is adjacent to $x$, so $x$ is a vertex of degree at least $k$
(\figref{staircase}.b).

Let $k=c\log n/\log \log n$, let $r=\sqrt{2/n}$, and let $X_1,\ldots,X_n$
be $n$ points drawn uniformly and independently from $[0,1]^2$.  Then the
area, $p$, of a step in a $(k,r)$-staircase is $p=(r/k)^2=2/nk^2$, so
\[
  \begin{aligned}
     \Pr\{\mbox{$X_1$ is a $(k,r)$-staircase} \mid X_1\in[r,1-r]^2\}
        & = \frac{(n-1)!}{(n-k-1)!} p^k (1-8/n)^{n-k-1} \\
        & \ge (1-k/n)^k (1-8/n)^{n} 2^k k^{-2k} \\
        & \ge k^{-2k} \enspace ,
  \end{aligned}
\]
for $n$ sufficiently large.
Thus, if $N$ is the number of staircases among $X_1,\ldots,X_n$, then
\[
  \EXP[N]\ge n(1-2r)^2k^{-2k}
    = \Omega(n^{1-2c-\epsilon})
      \rightarrow\infty \enspace ,
\]
provided that $c < 1/2$.

As before, we finish the proof using the second moment method.  Let $A$
denote the event that $\{X_1,X_2\}\not\subset[r,1-r]^2$ or that $X_2\in
X_1+[-r,r]^2$, and let $A^c$ denote the complement of $A$.  Let $Y_i$,
$i\in\{1,2\}$, denote the indicator variable that $X_i$ is a staircase.
Then, for sufficiently large $n$,
\[
  \begin{aligned}
    \frac{\EXP[Y_1Y_2]}{\EXP[Y_1]\EXP[Y_2]}
      & \le \frac{\EXP[Y_1Y_2]}{k^{-4k}} \\
      & = k^{4k} (\PROB\{A\}\EXP[Y_1Y_2|{A}]
             +\PROB\{A^c\}\EXP[Y_1Y_2|A^c]) \\
      & \le k^{4k} ((4r+4r^2) + \Pr\{A^c\}\EXP[Y_1Y_2|A^c]) \\
      & \le k^{4k} \left((4r+4r^2) + \frac{(n-2)!}{(n-2-2k)!}\left(\frac{1}{nk^2}\right)^{2k}(1-16/n)^{n-2-2k}\right) \\
      & \le k^{4k} \left((4r+4r^2) + n^{2k}\left(\frac{1}{nk^2}\right)^{2k}(1-16/n)^{n-2-2k}\right) \\
      & \le k^{4k}\left(4\sqrt{2/n}+ 8/n\right) + 1  \\
      & = 1 + O(n^{4c-1/2}) \enspace ,
  \end{aligned}
\]
so $\lim_{n\rightarrow\infty} \frac{\EXP[Y_1Y_2]}{\EXP[Y_1]\EXP[Y_2]} = 1$
for any $c<1/8$.
\end{proof}

\subsection{An upper bound}
\seclabel{yao-upper-bound}

Next we prove an upper bound on the maximum degree in a $(\pi/2)$-Yao graph.
The upper bound is based on the observation that the neighbours of a node
in a Yao graph are so-called minima.  Let $x_1,\ldots,x_n$ be a set of
points. We say that a point $x_i$ \emph{dominates} $x_j$ if the
$\mathrm{x}$- and $\mathrm{y}$-coordinate of $x_i$ are larger than the
$\mathrm{x}$- and $\mathrm{y}$-coordinate of $x_j$, respectively.  A point
$x$ is \emph{maximal} with respect to $x_1,\ldots,x_n$ if $x$ is not
dominated by any $x_i$.  A point $x$ is \emph{minimal} if $x$ does not
dominate any point $x_i$.

Before we can present the upper bound, we require a few preliminary results
about maxima and minima.  First, though, we recall a classic result 
obtained using Chernoff's bounding method \cite{c52}:

\begin{lem}\lemlabel{chernoff}
Let $Y_1,\ldots,Y_m$ be a sequence of independent $\{0,1\}$-valued
random variables, let $Y=\sum_{i=1}^m Y_i$, and let $\mu=\EXP[Y]$.
Then, for any, $\delta > 0$,
\[
   \Pr\{Y > (1+\delta)\mu\} 
     \le \left(\frac{e^{\delta}}{(1+\delta)^{(1+\delta)}}\right)^{\mu}
	\enspace .
\]
\end{lem}

The following result is already quite well-known.  We include a
proof sketch only for the sake of completeness.

\begin{lem}\lemlabel{maxima}
  Let $X_1,\ldots,X_m$ be a sequence of points drawn independently and
  uniformly from a rectangle $[a,b]\times[c,d]$ having area greater than
  0 and let $M$ be the number of maximal (respectively, minimal) points
  among $X_1,\ldots,X_m$.  Then, for any $\delta >0$,
  \begin{equation}
    \log m \le \EXP[M] \le \log m + 1
  \end{equation}
  and 
  \begin{equation}
    \PROB\{M>(1+\delta)\EXP[M]\} 
        \le \left(\frac{e^\delta}{(1+\delta)^{1+\delta}}\right)^{\log m}
		 \enspace .
  \end{equation}
\end{lem}

\begin{proof}
Sort the elements of $X_1,\ldots,X_m$ by decreasing
$\mathrm{x}$-coordinate, so that $X_i$ is maximal if and only if its
$\mathrm{y}$-coordinate is the maximum among the $y$-coordinates of
$X_1,\ldots,X_i$.  Let $Y_i=1$ if $X_i$ is maximal and $Y_i=0$ otherwise.
Obviously $\EXP[Y_i]=1/i$, so
\[
   \EXP[M] = \EXP\left[\sum_{i=1}^m Y_i\right] = \sum_{1=1}^m 1/i \enspace .
\]
The inequality $\log m \le \EXP[M]\le \log m + 1$ is then obtained by
bounding the above Harmonic sum using the integral $\int_{1}^{n}(1/x)dx$
(see, e.g., Cormen et al \cite[Appendix A.2]{clrs06}).

To prove the second part of the lemma, we use the fact that the random
variables $Y_1,\ldots,Y_m$ are independent \cite{d88,g78}.  The result
then follows immediately from \lemref{chernoff}.
\end{proof}

Unfortunately, the points we consider will not always be drawn from a
rectangle.  A \emph{$t$-shape} is a closed maximal subset of $\RR^2$ that
is bounded by the $\mathrm x$- and $\mathrm y$-axes and a $\mathrm
y$-monotone polygonal path consisting of at most $t$ edges.  See
\figref{t-shape}.a.

\begin{figure}
  \begin{center}
    \begin{tabular}{ccc}
      \includegraphics{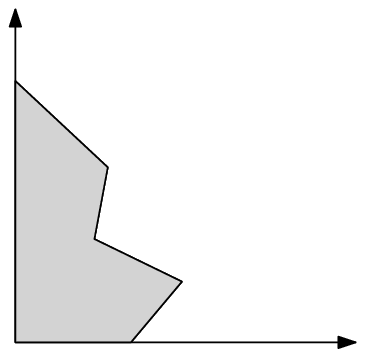} & 
      \includegraphics{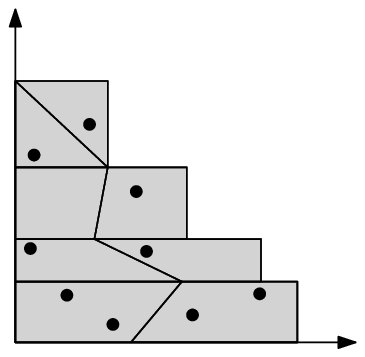} & 
      \includegraphics{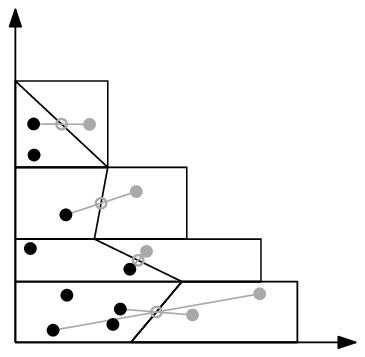} \\
      (a) & (b) & (c)
    \end{tabular}
  \end{center}
  \caption{(a)~a $t$-shape $S$, (b)~covering $S$ to obtain a shape $S'$ and
uniformly distributing points in $S'$, and (c)~reflecting the points in
$S'$ to obtain points uniformly distributed in $S$.}
  \figlabel{t-shape}
\end{figure}

\begin{lem}\lemlabel{t-shape-maxima}
  Let $X_1,\ldots,X_m$ be a sequence of points drawn independently and
  uniformly from a $t$-shape $S$ having area greater than
  0 and let $M$ be the number of minimal points
  among $X_1,\ldots,X_m$.  Then, for any $\delta >0$,
  \begin{equation}
    \EXP[M] \le 2t(\log m + 1)
  \end{equation}
  and 
  \begin{equation}
    \PROB\{M>(1+\delta)2t(\log m+1)\} 
        \le 2t\left(\frac{e^\delta}{(1+\delta)^{1+\delta}}\right)^{\log m} 
	\enspace .
  \end{equation}
\end{lem}

\begin{proof}
Cover $S$ with at most $t$ rectangles $R_1,\ldots,R_\ell$ whose total area
is twice the area of $S$, as shown in \figref{t-shape}.b.  Let
$S'=\bigcup_{i=1}^\ell R_i$ be the resulting subset of $\RR^2$.  Generate
points $Z=\{Z_1,\ldots,Z_m\}$ uniformly and independently in $S'$.  For each
point $Z_i$ in $R_j$, if $Z_i\in S$ then set $X_i=Z_i$.  Otherwise, set
$X_i$ to be the reflection of $Z_i$ through the center of $R_j$.  Observe
that $X_1,\ldots,X_m$ are uniformly distributed in $S$.  Furthermore, if
$X_i\in R_j$ is minimal with respect to $X_1,\ldots,X_m$, then $Z_i$ is either
maximal or minimal with respect to $Z \cap R_j$.

Therefore, if $M_j$ denotes the number of minimal elements of $X$ contained
in $R_j$, then, by the first part of \lemref{maxima}, $\EXP[M_j]\le
2(\log m+1)$ and 
\[
     \EXP[M] = \EXP\left[\sum_{j=1}^t M_j\right] \le 2t(\log m + 1)
     \enspace .
\]
By applying the second part of \lemref{maxima} $t$ times,
and using the union bound, we also obtain
\[
    \PROB\{M>(1+\delta)2t(\log m+1)\} 
        \le 2t\left(\frac{e^\delta}{(1+\delta)^{1+\delta}}\right)^{\log m} 
\]
as required.
\end{proof}

We now have all the tools required to prove our upper bound

\begin{thm}\thmlabel{yao-upper-bound}
For a random $(\pi/2)$-Yao graph defined on $n$ points drawn
independently from the uniform distribution on
$\D^2$, 
\[
\lim_{n\to\infty} \PROB \left\{ \hbox{\rm maximal degree}~> {c \log n \over
\log \log n} \right\} = 0
\]
for all $c > 4$.
\end{thm}

\begin{proof}
Let $X_1,\ldots,X_n$ be points uniformly and independently distributed in
$\D^2$ and let $G$ the the ($\pi/2$)-Yao graph of $X_1,\ldots,X_n$.
Let $\ell=\sqrt{d\log n/n}$.  We will first consider the edges of $G$ whose
length is at most $\ell$.  Consider the square $S=X_1+[0,\ell]^2$.  Let
$N$ denote the number of points of $X_2,\ldots,X_n$ contained in $S$.  Then
$\EXP[N] \le n\ell^2 = d\log n$ and, by \lemref{chernoff}, 
\[
   \Pr\{N > 2d\log n\} \le (e/4)^{d\log n} = n^{d(1-\log 4)} \enspace .
\]

Let $N'$ denote the number of points in $S$ that are neighbours of $X_1$
in the Yao graph.  Each such point is minimal with respect to the $N$
points of $X_2,\ldots,X_n$ contained in $S$.  Furthermore, $S\cap\D^2$ is
a $t$-shape, for $t\le 2$.  By the first part of \lemref{t-shape-maxima},
conditioned on $N \le 2d\log n$, the expected number of minimal points, and
hence the number of neighbours of $X_1$
in $S$ is small;
\[
   \EXP[N'|N \le 2d\log n]
        \le 4(\log(2d\log n) + 1) 
          = 4\log\log n + \Theta(1)\enspace .
\]
Define $v=\log(2d\log n)$ and
let $k=(c\log n)/(\log\log n)$.  By the second part of
\lemref{t-shape-maxima}, with $t=2$,
\[
  \begin{aligned}
    \PROB\left\{N'> k|N\le 2d\log n\right\}  
    & = \PROB\left\{N'> \frac{c\log n}{4(v+1)(\log\log n)}\cdot(4(v+1)) |N\le 2d\log n\right\} \\
    & \le4\left
         (\frac{\exp\left(\frac{c\log n}{4(v+1)(\log\log n)}-1\right)}
          {\left(\frac{c\log n}{4(v+1)(\log\log n)}\right)^
           \frac{c\log n}{4(v+1)(\log\log n)}}
         \right)^{\log(2d\log n)} \\
    & \le4\left
         (\frac{\exp\left(\frac{c\log n}{4(v+1)(\log\log n)}-1\right)}
          {\left(\frac{c\log n}{4(v+1)(\log\log n)}\right)^
           \frac{c\log n}{4(v+1)(\log\log n)}}
         \right)^{\log\log n} \\
    & \le4\left
         (\frac{\exp\left(\frac{c\log n}{4(v+1)(\log\log n)}\right)}
          {\left(\frac{c\log n}{4(v+1)(\log\log n)}\right)^
           \frac{c\log n}{4(v+1)(\log\log n)}}
         \right)^{\log\log n} \\
    & \le
         4\left(\frac{\exp\left(\frac{c\log n}{4(v+1)}\right)}
          {\left(\frac{c\log n}{4(v+1)(\log\log n)}\right)^{\frac{c\log
n}{4(v+1)}}} \right) \\
    & \le
         4\left(\frac{\exp\left(\frac{c\log n}{4(v+1)}\right)}
          {\left(\frac{c\log n}{4(v+1)}\right)^{\frac{c\log n}{4(v+1)}}} \right) \\
    & \le \frac{n^{o(1)}}{\Omega(n^{c/4-\epsilon})} \\
    & =  O(n^{-c/4 + \epsilon}) 
 \end{aligned} 
\]
for any $\epsilon > 0$.  Unconditioning, we obtain
\[
   \PROB\{N' > k\} = O(n^{-c/4+\epsilon} + n^{d(1-\log 4)}) \enspace .
\]
Let $G'$ be the subgraph of $G$ consisting only of edges of length at most
$\ell$ and let $D'$ denote the maximum degree of a vertex in $G'$.
Repeating the above argument $4n$ times and using the union bound gives
\[
   \PROB\{D' > k\} = O(n^{1-c/4+\epsilon} + n^{1+d(1-\log 4)})
\]

Finally, all that remains is to argue that $G$ has no edges of length
greater than $\ell$.  An edge of length at least $\ell$ defines an empty
region of area at least $\pi\ell^2/4$.  For $\ell < 1/2$, a portion of
this empty region whose area is at least $(\ell/2)^2$ is contained in
$\D^2$ (see \figref{empty}).  Therefore, the probability of there being
any edge of length greater than $\ell=\sqrt{d\log n/n}$ is at most
\[
 \begin{aligned}
   4n(1-\ell^2/4)^{n-2} 
    &  =  4n\left(1-\frac{d\log n}{4n}\right)^{n-2} \\
    & \le 4ne^{-(n-1)d\log n/4n} \\
    & \le 4n^{1-(n-1)d/4n} \\
    & = 4n^{1-(1-1/n)d/4} \\
 \end{aligned}
\]

\begin{figure}
  \begin{center}
    \includegraphics{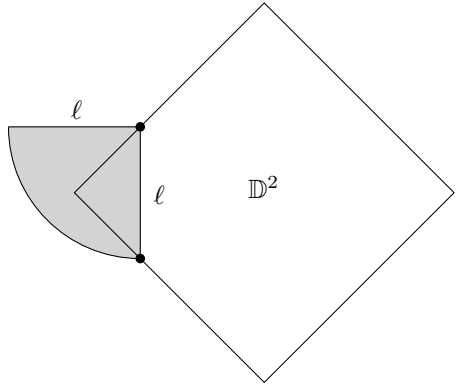}
  \end{center}
  \caption{An edge of length $\ell$ defines an empty subset of $\D^2$ whose
  area is at least $(\ell/2)^2$.}
  \figlabel{empty}
\end{figure}

At last, let $D$ be the maximum degree of any vertex in $G$.  Putting
everything together, we obtain
\[
   \Pr\{D > k\} = O(n^{1-c/4+\epsilon} 
                      + n^{1+d(1-\log 4)}
                      + n^{1-(1-1/n)d/4} )  \rightarrow 0
\]
for any
$c > 4$ and $d>\max\{1/(\log 4-1),4\}$.
\end{proof}

\subsection{Remarks}

\paragraph{Why $\D^2$?} 
The proof of \thmref{yao-upper-bound} actually shows that the
probability that $G$ has a vertex of degree more than $c\log n/\log\log n$
is $n^{-\Omega(c)}$.  The last step in the proof requires that any edge of
length $\ell$ defines portion of the support set of area $\Omega(\ell^2)$
that is empty of points.  This is true when the support set is $\D^2$
but not true when the support set is the unit square $[0,1]^2$.
Indeed, the proof breaks down for points drawn from the unit square,
since with probability $1/n$, some element, say $X_1$, simultaneously
has the minimum $\mathrm x$- and $\mathrm y$-coordinate.  In this case,
the expected degree of $X_1$ is equal to the expected number of minimal
elements among $X_2,\ldots,X_n$, which is, by \lemref{maxima}, $\Theta(\log n)$.

In a situation where points are uniformly distributed in the unit square,
the upper bound in \thmref{yao-upper-bound} holds if one considers only
the points whose distance from the boundary of the square is at least
$\sqrt{d\log n/n}$.

\paragraph{Smaller values of $\theta$.}
For any constant value of $\theta \le \pi /2$, the upper and lower bounds
of \thmref{yao-lower-bound} and \thmref{yao-upper-bound}
still hold.  The arguments are almost identical with the exception that
the definition of a staircase and of minima and maxima are modified to
take the value of $\theta$ into account.  Although the value of $\theta$
appears in the intermediate calculations, for any constant $\theta$,
the constants $c=1/8$ and $c=4$ in \thmref{yao-lower-bound}
and \thmref{yao-upper-bound} are unchanged.  However, as noted above,
to prove a version of \thmref{yao-upper-bound} the support set
must be rotated so that the difference in angle between any side of the
support set and $i\theta$, for $0\le i\le 2\pi/\theta$ is lower-bounded
by a constant.

\paragraph{Higher dimensions.}  
Yao graphs are also defined for point sets in $\RR^d$.  The lower bound
of \thmref{yao-lower-bound} can be extended to show that Yao graphs of
$n$ points uniformly and independently distributed in $[0,1]^d$ have
maximum degree $\Omega(\log n/\log\log n)$. Unfortunately, the proof of
the upper bound in \thmref{yao-upper-bound} does not continue to hold
in $\R^d$.

\bibliographystyle{plain}
\nocite{*}
\bibliography{proxdegree}

\end{document}